\documentclass{LT23auth}
\usepackage{graphicx}

\begin{document}

\begin{frontmatter}

\title{Structure and transport in multi-orbital Kondo systems}

\author[address1]{J. Kroha\thanksref{thank1}}, 
\author[address1]{S. Kirchner},
\author[address1]{G. Sellier},
\author[address1]{P. W\"olfle},
\author[address2]{D. Ehm}, 
\author[address2]{F. Reinert}, 
\author[address2]{S. H\"ufner},
\author[address3]{and C. Geibel}

\address[address1]{Universit\"at Karlsruhe, Institut f\"ur Theorie der
Kondensierten Materie, P.O.Box 6980, 76128 Karlsruhe, Germany}

\address[address2]{Universit\"at des Saarlandes, F.R.\ 2 -- 
Experimentalphysik, 
66041 Saarbr\"ucken, Germany}
 
\address[address3]{Max-Planck-Institut for Chemical Physics of Solids,
N\"othnitzer Str. 40, 01187 Dresden Germany}

\thanks[thank1]{Corr. Author. E-mail: kroha@tkm.physik.uni-karlsruhe.de}

\begin{abstract}
We consider Kondo impurity systems with multiple local orbitals,
such as rare earth ions in a metallic host or multi--level quantum
dots coupled to metallic leads. It is shown that the multiplet structure 
of the local orbitals leads to multiple Kondo peaks above the
Fermi energy $E_F$, and to ``shadow'' peaks below $E_F$. 
We use a slave boson mean field (MF) theory, which recovers the
strong coupling Fermi liquid fixed point, 
to calculate the Kondo peak positions,
widths, and heights analytically at $T=0$, 
and NCA calculations to fit the temperature
dependence of high--resolution photoemission spectra of Ce compounds.  
In addition, an approximate conductance quantization for transport
through multi--level quantum dots or single--atom transistors
in the Kondo regime due to
a generalized Friedel sum rule is demonstrated.
\end{abstract}

%
%
\end{frontmatter}

\section{Introduction}
In realistic Kondo impurity systems \cite{hewson93} the local magnetic moment
is often distributued over several local orbitals.
For example, in rare earth ions, embedded in a metallic host, 
the local moment is carried by the seven 4f orbitals
whose degeneracy is lifted by spin--orbit (SO) 
and crystal field (CF) splitting.
Small quantum dots (Qdot) containing few electrons are 
in general comprised of several single-particle levels 
and can be seen as
artificial atoms. Such systems, coupled to metallic leads, 
can be tuned to the Kondo regime as well,
if the Coulomb blockade is large enough to fix the electron
number on the Qdot to an odd number and, hence, the effective dot spin 
to 1/2. 
In rare--earth systems Kondo resonances have recently been
observed by high--resolution photoemission spectroscopy 
\cite{baer97,reinert01}, while the advances in nanotechnology
have made it possible to perform direct tunneling spectroscopy 
of the Kondo resonance in Qdots 
\cite{kouwenhoven98,goldhabergordon98,defranceschi02}.  

In this contribution 
we show that each of the low--lying single--particle orbitals
generates a many--particle Kondo resonance near the Fermi energy
$E_F$, and we calculate their positions, widths and heights at
temperature $T=0$. The results of a Non--Crossing Approximation (NCA)
calculation at finite $T$ are compared to high--resolution 
photoemission spectra of the rare--earth Kondo system CeSi$_2$. 
Moreover, an approximate conductance
quantization in multi--orbital Kondo Qdots is predicted
originating from a generalized Friedel sum rule \cite{kirchner01}. 

The systems discussed above can be described by the multi--orbital Anderson
impurity Hamiltonian
\begin{eqnarray}
H &=& H_{{\rm kin}} + \sum_{m\sigma} \varepsilon _{d m}
d_{m\sigma}^{\dag} d_{m\sigma}^{\phantom{\dag}}
+\sum_{p m\sigma} [V_{mp} d_{m\sigma}^{\dag} c_{p\sigma} +h.c.] \nonumber\\
&+&\frac{U}{2} \sum _{(m\sigma)\neq (m'\sigma')} 
d_{m\sigma}^{\dag} d_{m\sigma}^{\phantom{\dag}}\ d_{m'\sigma'}^{\dag} 
d_{m'\sigma'}^{\phantom{\dag}},
\label{hamilton}
\end{eqnarray} 
where
$H_{{\rm kin}} = \sum_{p\sigma} 
\varepsilon _p c_{p\sigma}^{\dag} c_{p\sigma}^{\phantom{\dag}}$
describes the conduction electron band, and $d_{m\sigma}^{\dag}$,
creates an electron with spin $\sigma$ in the low--lying single--particle
level $\varepsilon _{dm} < 0$, $m=0,\dots,M-1$, with
$\varepsilon _{d0}$ the (noninteracting) ground state. 
Electrons in any of the local
orbitals have a Coulomb repulsion $U$ and are
coupled to the conduction band via the hybridization matrix
elements $V_{mp}$. 
For later use we indroduce the effective couplings 
$\Gamma _{mm'} = \pi \sum_p V_{mp} V_{pm'}^* A_{p\sigma}(0)$, 
with $A_{p\sigma}(\omega)$
the conduction electron spectral function and 
$N(\omega)=\sum _p A_{p\sigma}(\omega)$ the density of states per spin. 
In the following we will assume the system to be in the
Kondo regime, $\Gamma _{mm} / |\varepsilon _{dm}|\ll 1$, $U\to \infty$, and
for the expansions leading to Eqs.\ (\ref{Gf}), (\ref{position}), and
(\ref{width1}) below, 
$\Gamma _{mm}/(\varepsilon _{dm} -\varepsilon _{d0}) \ll 1$.

\section{Spectral features}

In 2nd order perturbation theory, the spin scattering amplitude
of the local spin shows log divergences at the energies 
$(\varepsilon _{dm}-\varepsilon _{d0})\ll |\varepsilon _{d0}|$ (*)
and  $-(\varepsilon _{dm}-\varepsilon _{d0})$ (**).
These instabilities can be understood as due to spin fluctuations
involving a transition of the local electron from the
ground state $\varepsilon _{d0}$ to an excited state 
$\varepsilon _{dm}$ ($m>0$) (*) or vice versa (**) 
\cite{reinert01,kirchner02}.
The positions of these resonances at $T=0$ can be estimated 
analytically from a slave boson MF theory \cite{coleman84}
for Eq.\ (\ref{hamilton}), since this approach recovers the 
strong coupling Fermi liquid fixed point \cite{hewson93}
of the model. The numerical evaluations are done within
the NCA \cite{reinert01}, which correctly describes the
life-times of the excited resonances and the temperature dependence  
for $T \stackrel{>}{\sim} T_K$ \cite{else}. 

In order to implement the effective restriction of
no double occupancy of {\it all} orbitals, enforced by the strong 
Coulomb repulsion $U$, the local electron operator is represented 
in terms of pseudofermions $f_{m\sigma}$ and slave bosons $b$
as $d_{m\sigma}^{\dag}=f_{m\sigma}^{\dag} b$, with the
operator constraint $\hat Q := \sum _{m\sigma} 
f_{m\sigma}^{\dag} f_{m\sigma}^{\phantom{\dag}} + b^{\dag} b \equiv 1$.  
In a functional integral approach, the latter is imposed by the
functional Kronecker delta 
\begin{eqnarray}
\Delta (Q-1) = \int _{-\pi T}^{\pi T} \frac{{\rm d}\lambda}{2\pi T}\ {\rm e}
^{-i\beta\lambda (Q-1)}\ .
\label{constraint}
\end{eqnarray} 
Introducing mean field and fluctuating parts $r$, $a$  
for the Bose field, $b = r + a$, $b^{\dag} = r + a^{\dag}$
and $\lambda _0$, $\tilde \lambda$ for the
constraint field, $\lambda = -i \lambda _0 +\tilde\lambda$, 
Eq.\ (\ref{hamilton}) with Eq.\ (\ref{constraint}) takes the form
\cite{hewson93}
\begin{eqnarray}
H &=& H_{{\rm kin}} + \sum_{m\sigma} \tilde \varepsilon _{d m}
f_{m\sigma}^{\dag} f_{m\sigma}^{\phantom{\dag}}
+\sum_{p m\sigma} [\tilde V_{mp} f_{m\sigma}^{\dag} c_{p\sigma} +h.c.] 
\nonumber\\
&+& (\tilde \varepsilon _{d0} - \varepsilon _{d0} )[r^2 + a^{\dag} a -1]
+H_{{\rm int}}
\end{eqnarray} 
\begin{eqnarray}
H_{{\rm int}}&=&
\sum_{p m\sigma} [V_{mp} a f_{m\sigma}^{\dag} c_{p\sigma} +h.c.]
+ (\tilde \varepsilon _{d0} - \varepsilon _{d0} ) r [a^{\dag} + a ]
\nonumber\\
&+& \tilde\lambda 
\Bigl[ a^{\dag} a + r^2 + r(a^{\dag} + a) + 
\sum_{m\sigma} f_{m\sigma}^{\dag} f_{m\sigma}^{\phantom{\dag}} -1 \Bigr] \ .  
\end{eqnarray} 
\begin{figure}[t]
\begin{center}\leavevmode
\includegraphics[width=0.8\linewidth]{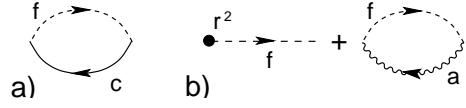}
\caption{Diagrammatic representation of a) the fluctuating Bose
selfenergy and b) the local physical electron Green's function. 
The solid circle represents the modulus squared of the boson MF.}
\label{fig1}\end{center}
\end{figure}
\noindent
Here we have defined renormalized local levels $\tilde \varepsilon _{dm}$
such that
$\lambda _0 = \tilde \varepsilon _{d0} - \varepsilon _{d0}$ and
$\tilde \varepsilon _{dm} = \varepsilon _{dm} + \lambda _0 $ 
and effective hybridizations $\tilde V_{mp} = r V_{mp}$,
$\tilde \Gamma_{mm} = r^2 \Gamma_{mm}$.
The mean field solutions $r$, $\tilde\varepsilon _{d0}$ are then 
obtained by minimizing the free energy, which is equivalent to
setting the terms linear in the fluctuating fields $a$, $a^{\dag}$,
$\tilde\lambda $ equal to zero,
\begin{eqnarray}
r^2 + \langle a^{\dag} a \rangle 
+ \sum_{m\sigma} 
\langle f_{m\sigma}^{\dag} f_{m\sigma}^{\phantom{\dag}}\rangle -1 &=& 0 
\label{MF1}\\
(\tilde \varepsilon _{d0} - \varepsilon _{d0} ) r  
+\sum_{p m\sigma} V_{mp} \langle f_{m\sigma}^{\dag} c_{p\sigma} \rangle 
&=& 0 \ .
\label{MF2}
\end{eqnarray}
The expectation values in Eqs.\ (\ref{MF1}), (\ref{MF2}) are 
calculated from the
Fourier components of the pseudofermion, the mixed
fermion--conduction electron, and the fluctuating boson Green's functions
$G_{f\ m m'\sigma}(t) = -i \langle \hat T f_{m\sigma}^{\phantom{\dag}}(t) 
f_{m' \sigma}^{\dag}(0)\rangle $, 
$G_{cf\ pm\sigma}(t) = -i \langle \hat T c_{p\sigma}^{\phantom{\dag}}(t) 
f_{m\sigma}^{\dag}(0)\rangle $, 
$G_a(t) = -i \langle \hat T a^{\phantom{\dag}}(t) a^{\dag}(0)\rangle $.
Calculating the $f$ selfenergies for a flat band and keeping
only the leading terms in the effective hybridizations, they are
\cite{kirchner02,kirchner02a},
\begin{eqnarray}
G_{f\ mm\sigma}(i\omega)&=&
\bigl[ i\omega - \tilde\varepsilon _{dm} - i\tilde\Gamma _{m m}  
\bigr] ^{-1}
\label{Gf}\\
G_{cf\ pm \sigma} (i\omega)&=&
\tilde V_{mp}^* \ G_{c p\sigma}(i\omega) \ G_{f\ mm\sigma}(i\omega)\\  
G_a (i\nu ) &=& 
\bigl[ i\nu - (\tilde\varepsilon _{d0} - \varepsilon _{d0} )
- \Sigma _a (i\nu)  \bigr] ^{-1}\ .
\end{eqnarray}
As seen from Eq.\ (\ref{ad}) below, the low-energy 
peak in each of the pseudofermion spectral functions 
$G_{f\, mm}$ is carried over to a quasiparticle peak in
the physical impurity electron spectral function, i.e. to a Kondo resonance. 
Hence, the width $\tilde\Gamma _{mm}$ of $G_{f\, mm}$ {\it defines}
the Kondo temperature of the $m$th Kondo peak, 
$\tilde\Gamma _{mm} = T_{K\, m}$. It is to be determined 
from the MF solution below. The excitied resonances have 
an additional width due to inelastic relaxation \cite{else}.
The selfenergy of the fluctuating Bose field appearing in
Eq.\ (\ref{MF1}) is given by (Fig.\ \ref{fig1} a)), 
\begin{eqnarray}
&{\rm Im}&\Sigma _a (\omega -i0) = \\
&\frac{2}{\pi}&\sum _m \Gamma _{mm} 
\Bigl[
\arctan \Bigl( \frac{\omega -\tilde \varepsilon _{dm}}{T_{K\, m}} \Bigr) +
\arctan \Bigl( \frac{\tilde \varepsilon _{dm}}{T_{K\, m}} \Bigr)
\Bigr] \ . \nonumber
\end{eqnarray}
It is worth noting that Eq.\ (\ref{MF1}) just represents the
Friedel sum rule, when the expectation value $\langle a^{\dag}a\rangle$
of the fluctuating boson density is included 
\cite{kirchner02,kirchner02a}. Therefore, it
describes the correct shift of the Kondo resonances w.r.t. $E_F$
due to potential scattering. 
We now use the relations $\tilde\varepsilon _{d0} < T_{K\, 0}$ and
$\tilde\varepsilon _{dm} > T_{K\, m},\ m=1,\dots,M-1$, where the
former will be justified below by the MF solution in the Kondo regime, 
and the latter is true for sufficiently large splitting of the 
local levels. The positions of the Kondo peaks and the 
renormalized $T_K$ 
are obtained from Eqs.\ (\ref{MF2}), (\ref{MF1}),
respectively, as
\begin{eqnarray}
\tilde\varepsilon _{d0} &=& \frac{\Gamma_{00}}{-\pi\varepsilon_{d0}}
\Bigr(1+ \sum _{m=1}^{M-1}  
\frac{-2 \varepsilon_{d0}\ T_{K\, m}}
{\Gamma _{00}(\varepsilon _{dm}-\varepsilon _{d0})}\Bigl)\ T_{K\, 0}
\label{position}\\
\tilde\varepsilon _{dm} &=& \tilde\varepsilon_{d0} + 
(\varepsilon _{dm} -\varepsilon _{d0})\\
T_{K\, 0} &=&  \prod _{m=1}^{M-1} 
\bigl( \frac{D}{\varepsilon _{dm}-\varepsilon _{d0}}\bigr) 
^{\frac{\Gamma _{mm}}{\Gamma _{00}}}
\ D\ {\rm e} ^{-\frac{\pi|\varepsilon _{d0}|}{2\Gamma_{00}}} \ .
\label{width1}
\end{eqnarray}
\begin{figure}[t]
\begin{center}\leavevmode
\includegraphics[width=0.8\linewidth]{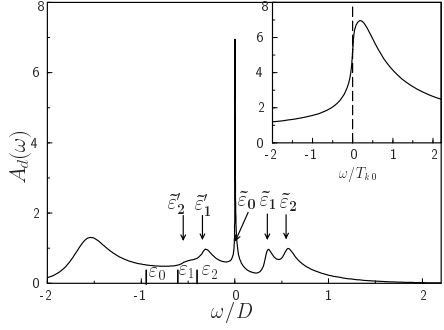}
\caption{Local electron spectrum of the Hamiltonian Eq.\ (\ref{hamilton}),
NCA calculation, $M=3$. $\varepsilon _{d0,1,2} = -0.95D, -0.6D, -0.4D$,
$\Gamma _{00} =\Gamma _{11} =\Gamma _{22} =
 0.2D$;  $T_{K0}= 3\cdot 10^{-3}D$; $T=5\cdot 10^{-5}D$;
$D$=half band width.}
\label{fig2}\end{center}
\end{figure}
\noindent
Finally, we obtain the physical electron Green's function
by evaluating the diagrams shown in Fig.\ \ref{fig1}b). At $T=0$ its
imaginary part reads
\begin{eqnarray}
&A_{d\sigma }& (\omega ) \simeq 
\sum _m \frac{T_{K\, m}}{\Gamma_{mm}}  
{\rm Im} G_{f\, mm\sigma} (\omega -i0) \nonumber \\
&&-\bigl[\Theta (\tilde\varepsilon _{d0}-\omega) -\frac{1}{2}\bigr] 
\ {\rm Im} \Bigl[ -\omega +\tilde\varepsilon _{d0}+\varepsilon _{d0}
\label{ad} \\
&&\hspace*{0.6cm}  - \sum_m \frac{\Gamma _{mm}}{\pi}
\ln \frac{(\omega +\tilde\varepsilon _{dm}-
\tilde\varepsilon _{d0})^2 + T_{K\, m}^2}
         {\tilde\varepsilon _{dm}^2 + T_{K\, m}^2} \nonumber \\ 
&&\hspace*{0.6cm}
-i {\rm Im} \Sigma _a (\omega-i0) -i\ {\rm sgn} \omega\ T_{K\, 0}
\Bigr]^{-1} \ . \nonumber
\end{eqnarray}
It is seen that the first line of Eq.\ (\ref{ad}) represents 
multiple Kondo resonances above $E_F$ with the positions and 
weights given by Eqs.\ (\ref{position})--(\ref{width1}).
The remainder of Eq.\ (\ref{ad}) describes the structure below $E_F$.
The single--particle resonance ($\varepsilon _{d0}$) is strongly
renormalized downward due to the logarithmic terms.
The maxima of the log terms indicate 
weak peaks or often merely shoulders at the energies
$\tilde\varepsilon _{dm}' = -(\tilde\varepsilon _{dm}-
\tilde\varepsilon _{d0})$,
i.e. at mirrorimaged positions w.r.t. $\tilde\varepsilon _{d0}$. 
In contrast to the resonances above $E_F$, these mirrored peaks 
do not correspond to quasiparticle resonances, since there is
no pole in $A_{d\sigma}$ at these energies. Physically, they
originate from spin fluctuations involving a virtual transition
from an excited state ($m\geq 1$) to the ground state ($m=0$).
This is only a virtual process, as the $m\geq 1$ states are
not thermally occupied at $T=0$ \cite{kirchner02}. 
With increasing $T$, $0<T<T_{K\, m}$, the peaks below $E_F$
grow, as the $m\geq 1$ states become thermally occupied.
The multiple peak structure
discussed above is shown in Fig.\ \ref{fig2}.
Fig.\ \ref{fig3} shows fits of the NCA calculation of $A_d$ to 
photoemission spectra of the Kondo system CeSi$_2$ ($M=7$)  
\cite{note}. The experimental data show the appearance of 
low--energy peaks above $E_F$ at elevated $T$, in 
quantitative agreement with the $T$ dependence expected 
theoretically from multiple Kondo resonances.

\begin{figure}[h]
\begin{center}\leavevmode
\includegraphics[width=0.75\linewidth,angle=0]{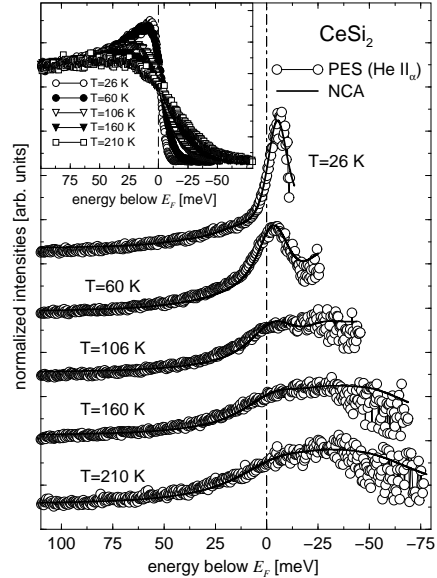}
\caption{Low-energy high resolution photoemission spectra of CeSi$_2$.
Open circles: Experiment; solid lines: NCA. Model parameters determined from 
fits to the theory:
$\varepsilon _{d0}$=-1.35 eV, SO splitting: 270 meV,
CF splitting: 25 meV (lower quartet), 48 meV (upper sextet);
$D$ = 3.7 eV; $T_{K0}\approx 35$ K.
The inset shows raw data before division by the Fermi-Dirac
distribution \cite{reinert01}. 
}
\label{fig3}\end{center}
\end{figure}
\noindent

\section{Approximate conductance quantization}

We now show that in a symmetrical multi--level Qdot or a single--atom 
transistor (SAT) \cite{ralph02}
at low $T$ the linear response conductance $G={\rm d}I/{\rm d V}(V=0)$ is 
{\it approximately} 
quantized at $2e^2/h$ \cite{kirchner01}, if the Qdot or SAT  
is in the Kondo regime,
even though there may be several transmission channels, $m=0,\dots,M-1$, 
and the lead--to--dot coupling matrix elements $V_{mp}$ are random.
Such systems are described by the
Hamiltonian Eq.\ (\ref{hamilton}), where, in addition, the 
conduction electron operators carry an index $\alpha = L,R$ 
indicating the left or right lead, and, hence the 
kinetic energy term reads, 
$H_{{\rm kin}} = \sum_{p\sigma\alpha} 
[\varepsilon _p -\mu _{\alpha}]
c_{p\sigma\alpha}^{\dag} c_{p\sigma\alpha}^{\phantom{\dag}}$,
with $\mu_{L/R} =0, V$ the chemical potentials in the left/right lead.

The current through a left/right symmetrical device is \cite{wingreen.92}
\begin{eqnarray}
I=\frac{e}{h} \sum _{\sigma} \int d\omega 
\Bigl[f(\omega )-
      f(\omega +\frac{eV}{\hbar} )\Bigr] {\rm Im}\, {\rm tr}\,   
[\Gamma {G}_{d\sigma}(\omega )] , \nonumber \\
\label{eq:current}
\end{eqnarray}
where ${G}_{d\sigma}$ is the advanced local-orbital Green's function
and $\Gamma$ the lead--dot couplings $\Gamma_{mm'}^L+\Gamma_{mm'}^R$.
$G_{d\sigma}$ and $\Gamma$ are matrices in the space of local 
orbitals,
\begin{eqnarray}
\left[ G^{-1}_d \right] _{mm\sigma}(\omega)&=& 
\omega-\varepsilon_{d,1}-i\Gamma_{mm}-\Sigma_{mm}(\omega)
\label{gdiag}\\
\left[ G^{-1}_d \right]  _{mm'\sigma}(\omega)&=& 
-i\Gamma_{mm'}-\Sigma_{mm'}(\omega )\ ,\ \  m\neq m', 
\label{goffdiag}
\end{eqnarray}
where $\Sigma_{mm'}(\omega)$ is the selfenergy due to intra--dot
interactions. 
Since the ground state is a spin singlet \cite{hewson93}, 
the quantum dot is at $T\ll T_K\equiv {\rm min}[T_{K\, m}]$
a pure potential scatterer for electrons traversing the system, and the
following Fermi liquid relations hold \cite{luttinger.60}, 
\begin{eqnarray}
\label{eq:fl}
&\phantom{=}&
\Sigma ''_m (\omega ) = \frac{ (\hbar\omega) ^2 + (\pi k_BT)^2 }{k_BT_K} 
\hspace*{0.6cm}\omega, T < T_K\\
&\phantom{=}&
\int _{-\infty}^{0} d\omega\; {\rm tr}\Bigl\{ 
   \frac{\partial \Sigma (\omega)}{\partial \omega} \cdot 
   {G}_{d\sigma}(\omega) \Bigr\}=0
\label{eq:lutt}
\end{eqnarray}
The averaged electron number in the dot
per spin, $n_{d,\sigma}$, 
can now be evaluated using the general relation   
$\frac{d}{d\omega}{\rm ln}({G_d}^{-1}) = 
(1 - \frac{d\Sigma}{d\omega})\cdot {G_d}$
and the Luttinger theorem Eq.~(\ref{eq:lutt}),
\begin{eqnarray}
n_{d\sigma} = {\rm Im} \int _{-\infty}^0 \frac{d\omega}{\pi}
{\rm tr}\; {G_{d\sigma}(\omega )} =
\frac{\rm Im}{\pi}  
\Bigl[{\rm tr}\{ {\rm ln}\; {G}_{d\sigma}(\omega )^{-1}\} 
\Bigr] _{-\infty}^{0}
\label{eq:friedel} \nonumber
\end{eqnarray}
This is a statement of the Friedel sum rule
$n_{d\sigma} = \frac{1}{\pi} \sum _{m} \delta _{m\sigma} (0) $, since 
the scattering phase shift at the Fermi level in channel $m$
is $\delta _{m\sigma}(0) =
{\rm arg} [\Gamma \cdot {G}_{d\sigma}(0)]_{mm}$.
It may be re-expressed, using 
${\rm tr}\;{\rm ln}\;{G_{d\sigma}}^{-1} = 
{\rm ln}\;{\det}\;{G_{d\sigma}}^{-1}$, as
\begin{eqnarray}
n_{d\sigma} =  \frac{1}{\pi}{\rm arccot}
 \Biggl[ \frac{{\rm Re}\;{\rm det}\;{G}_{d\sigma}(0)^{-1}}
              {{\rm Im}\;{\rm det}\;{G}_{d\sigma}(0)^{-1}}\Biggr]\; .
\end{eqnarray}
The scattering T-matrix of the device, 
$\Gamma \cdot {G} _{d\sigma}$, which determines the conductance $G=dI/dV$
of the system via Eq.\ (\ref{eq:current}), is now evaluated
by expressing the inverse matrix Eqs.\ (\ref{gdiag}), (\ref{goffdiag})  
in terms of its determinant, and, using the Fermi liquid property
Eq.~(\ref{eq:fl}), we obtain at the Fermi energy ($\omega =0$, $T\ll T_K$)
for $M=2$,
\begin{eqnarray}
\label{eq:unitarity}
&&{\rm Im}\; {\rm tr}\; (\Gamma \cdot {G}_{\sigma}(0)) = 
{\rm sin}^2(\pi n_{d\sigma}) + \\ 
&&  {\rm sin}(2\pi n_{d\sigma}) 
\frac{{\rm Re}[{\rm det}(i\Gamma -\Sigma '(0))]}
{\Gamma_{11}(\varepsilon_{d,2}+\Sigma _{22}'(0))+
 \Gamma_{22}(\varepsilon_{d,1}+\Sigma _{11}'(0))}\; .
\nonumber
\end{eqnarray}
If the transition amplitudes $V_{mp}$ are independent of the
lead channels $p$, it follows directly from the definition of 
$\Gamma _{mm'}$
that the term $\propto {\rm sin}(2\pi n_{d\sigma})$ cancels. 
Eq.~(\ref{eq:unitarity}) is an exact result, valid for arbitrary
microscopic parameters $\Gamma _{mm'}$, 
$\varepsilon _{dm}$, $U$, and $n_{d\sigma}$.
It is the generalization of the well-known unitarity rule of the single-level 
Anderson impurity problem to the case of several impurity levels 
\cite{kirchner01}. If there is 
at least one of the local levels significantly below the Fermi level
($\varepsilon _{d0} < 0$, $|\varepsilon _{d0}| /\Gamma _{mm'} < 1$), 
a strong Coulomb repulsion $U$ enforces $n_{d\sigma} \approx 1/2 $, implying
via Eqs.~(\ref{eq:unitarity}), (\ref{eq:current}) a conductance 
close to the conductance unit, i.e.~$dI/dV \approx 2 e^2 /h$
(the factor 2 reflects spin summation). 
%
%
\begin{ack}
Stimulating discussions with
C. Cuevas, A. Rosch, and E. Scheer are gratefully acknowledged.
This work was supported by CFN and SFB195
of the Deutsche Forschungsgemeinschaft.
\end{ack}

%
%

\end{document}